\documentstyle[aps,preprint]{revtex}
\def\prl#1#2#3{{ Phys. Rev. Lett.,} {\bf #1}, #2 (#3)}

\def\pla#1#2#3{Phys. Letts. A {\bf #1}, #2 (#3)}
\def\pre#1#2#3{Phys. Rev. E, {\bf #1}, #2 (#3)}
\def\pra#1#2#3{Phys. Rev. A, {\bf #1}, #2 (#3)}

\def\physd#1#2#3{Physica D {\bf #1}, #2 (#3)}

\def\etc{$ etc.$~}

\def\etl{$et.~al.$~}
\def\MLE{$\Lambda$}

\def\beq{\begin{equation}}
\def\bc{\begin{center}}
\def\ec{\end{center}}
\def\eqn{\end{equation}}
\begin{document}
\title{Strange Nonchaotic Attractors in the Quasiperiodically forced
Logistic Map}
\author{Awadhesh Prasad, Vishal Mehra and Ramakrishna Ramaswamy} 
\address{School of Physical Sciences\\ Jawaharlal Nehru University, New Delhi
110 067, INDIA}
\date{\today}
\maketitle
\begin{abstract}
Different mechanisms for the creation of strange non-chaotic dynamics
in the quasiperiodically forced logistic map are studied. These routes
to strange nonchaos are characterised through the behavior of the
largest nontrivial Lyapunov exponent, as well as through the
characteristic distributions of finite time Lyapunov exponents.
Strange nonchaotic attractors can be created at a saddle--node
bifurcation when the dynamics shows Type-I intermittency; this {\it
intermittent}~ transition, which is studied in detail, is characterised
through scaling exponents.  Band-merging crises through which dynamics
remains nonchaotic are also studied, and correspondence is made with
analogous behavior in the unforced logistic map. Robustness of these
phenomena with respect to additive noise is investigated.

\end{abstract}

\newpage
\section{INTRODUCTION}

Of the diverse structures that are found in nonlinear dynamical
systems, strange nonchaotic attractors (SNAs) are among the more
exotic. These were first described by Grebogi \etl \cite{gopy} in a the
context of quasiperiodically forced systems, and are characterised by
having a geometrically strange structure (usually a fractal, which is
everywhere single valued but piecewise nondifferentiable) with
``nonchaotic'' dynamics: the largest nontrivial Lyapunov exponent
$\Lambda$ is negative, and nearby orbits do not diverge from each other
exponentially. Since they were first described \cite{gopy}, a number of
characteristics of SNAs have been studied theoretically
\cite{gory,rboag,kpf,hh,k,nk,yl,blowout,pf,sfkp,lai,pm} and
experimentally \cite{ditto,bulsara,newexp}.

Strange nonchaotic dynamics usually occurs in the vicinity of strange
chaotic behavior and periodic or quasiperiodic (nonstrange, nonchaotic)
behavior. The different mechanisms through which SNAs are created,
either from regular or from chaotic motion, and the mechanisms through
which they disappear, either into regular or chaotic motion---the birth
and death of strange nonchaotic attractors, so to speak---is a topic of
considerable current interest.

In this paper, we address this question in the context of a typical
dynamical system with quasiperiodic forcing and discuss a number of
transitions in such systems. These include, apart from the transition
from quasiperiodic or chaotic motion to strange nonchaotic dynamics,
transitions between different SNAs.

We study the quasiperiodically forced logistic map which is given by
the equations
\begin{eqnarray}
x_{n+1} &=& \alpha (1 + \epsilon \cos (2 \pi \phi_n))~x_n~ ( 1 - x_n)
\nonumber \\ 
\phi_{n+1} &=& \phi_n + \omega ~~~~~({\rm mod ~} 1),
\label{qlm}
\end{eqnarray}\noindent
where $x \in \hbox{R}^1$, $\phi \in \hbox{S}^1$, $\omega=
(\sqrt{5}-1)/2$ is the irrational driving frequency and $\epsilon$
represents the forcing amplitude. With quasiperiodic driving the
periodic attractors of the logistic map become quasiperiodic
attractors, and following the standard notation
\cite{hh}, we denote a torus attractor of period $n$ in R$^1 \times
$S$^1$ as $n$-T.

Four mechanisms or scenarios for the creation of SNAs in
quasiperiodically driven systems have been advanced. In the absence of
quasiperiodic forcing, the corresponding systems show the by now
standard scenarios for the route to chaotic or aperiodic behavior,
including quasiperiodicity, period doubling and tangent bifurcations,
intermittency, crises, and band merging or reverse bifurcations
\cite{ott}. There are parallels to several of these in the different
routes to SNAs.

\begin{itemize}
\item
Heagy and Hammel \cite{hh} identified the birth of a SNA with the
collision between a period-doubled torus and its unstable parent. This
mechanism requires that a period doubling bifurcation occur, after
which the stable torus attractor gets progressively more ``wrinkled''
as the parameters in the system change, {\it i.e.} $x(\phi)$ becomes
more and more oscillatory. Following a collision with the unstable
parent torus, at an analogue of the attractor--merging crisis that
occurs in chaotic systems \cite{gory}, the SNA is born.  \MLE~ remains
negative throughout the collision process. This mechanism, which we
denote by HH in the remainder of this paper, has been seen both for a
period 2 torus as well as for the period 4 torus; the quasiperiodic
forcing drives the system into chaos well before the infinite sequences
of period doubling can occur.

\item
The ``fractalization'' route for the creation of SNAs has been
described by Kaneko \cite{k,nk}. A torus gets increasingly wrinkled and
transforms into a SNA without any interaction with a nearby unstable
periodic orbit.  This route to SNA (and eventually to chaos) has also
been observed in higher dimensional systems \cite{sfkp}.

\item
In the intermittency scenario for the formation of SNAs \cite{pmr}, as
a function of driving parameter a strange attractor disappears and is
eventually replaced by a 1-frequency torus through an analogue of the
saddle-node bifurcation.  In the vicinity of this crisis--like
phenomenon \cite{gory} the attractor is strange and nonchaotic. We have
shown that the dynamics at this transition shows scaling behavior
characteristic of Type I intermittency \cite{pm}, and the signature of
the transition is an abrupt and characteristic change in the variation
of \MLE. The intermittent route is general one which can be found in
other systems as well \cite{ring}.

\item
Yal\c{c}inkaya and Lai \cite{yl} identify the birth of SNAs with a
blowout bifurcation \cite{blowout}, when a torus loses transverse
stability. The Lyapunov exponent also has a characteristic dependence
on parameter in this case.

\end{itemize}

SNAs can be quantitatively characterised by a variety of methods,
including the estimation of the Lyapunov exponents and the fractal
dimension \cite{gopy,rboag,dgo}, spectral properties \cite{rboag,nk},
examination of the time-series \cite{sfkp,kn}. The geometric
strangeness of the attractor can be measured through indices such as
the phase--sensitivity exponent \cite{pf}, while the chaoticity
properties can be studied by examining the finite time Lyapunov
exponents \cite{pf,pmr}.

A number of transitions in this system are investigated wherein three
of the above mechanisms for the creation of SNAs are known to be
operative (blowout bifurcations cannot occur), notably the processes
whereby SNAs are created from torus (T) attractors
\begin{eqnarray}
\ldots n~\hbox{T} \leftrightarrow n~\hbox{ band SNAs}\leftrightarrow 
\ldots \\
2^n~\hbox{T} \leftrightarrow 2^{n-1}~\hbox{ band SNAs}\leftrightarrow
\ldots 
\end{eqnarray}
or others, such as from SNA to chaotic attractors (C) or from a
$k$--band SNAs to a $k/2$--band SNA,
\begin{eqnarray}
\ldots n~\hbox{ band SNAs}\leftrightarrow n~\hbox{ band
C}\leftrightarrow.... \\
\ldots \leftrightarrow k~\hbox{ band SNAs}\leftrightarrow k/2~
\hbox{band SNAs}\leftrightarrow k/2~\hbox{ band C}\leftrightarrow \ldots
\end{eqnarray}
The latter class of bifurcations (for $k$ a power of 2) are studied
here through analysis of the Lyapunov exponents and their
distributions. We also explore the effect of additive noise.

This paper is organized as follows. In Section~II, we describe the
``phase--diagram'' for the forced logistic map: the regions
corresponding to the different dynamical behavior that obtain are
delineated as a function of parameters. This phase diagram is canonical
for dynamical systems such as Eq.~(\ref{qlm}), namely unimodal maps
that are parametrically driven, and generalizes the usual bifurcation
diagram that is obtained in the absence of forcing. Transitions to SNAs
and the characteristic behavior of the Lyapunov exponents are discussed
in Section~III, where we also discuss other transformations that are
undergone by SNAs, such as the analogue of band--merging. The creation
of SNAs is often accompanied by intermittent dynamics, and this can be
quantitatively described in terms of scaling exponents at these
transitions. These results are also given in Section~III, where we
discuss, in addition, the effects of additive noise. This is followed
by a summary in Section~IV.

\section{Forced Logistic Map: Phase diagram}

The quasiperiodically forced logistic map \cite{hh} is particularly
convenient for study since the phenomenology is smoothly related to
that of the logistic map in the limit of $\epsilon \to 0$. Since $x$
and $\phi$ are uncoupled in this limit, the period--$k$ orbits of the
logistic map are converted into $k$--frequency tori, and the chaotic
attractors in the logistic map appear as chaotic band attractors of the
two-dimensional map, Eq.~(\ref{qlm}).

The region of interest in the phase space is $0 \le x \le 1$, $0 \le
\phi \le 1$. For $\epsilon \ne 0$, it is clear that motion will
remain bounded in this region so long as $\alpha (1 + \epsilon \cos (2
\pi \phi_n))
\in [0,4]$. Thus for any $\alpha \le 4$, the largest value of $\epsilon$
allowed is $4/\alpha -1$.  We therefore redefine the driving parameter
as $\epsilon^{\prime} =\epsilon/(4/\alpha -1 )$ and study the system
for $0 \le \epsilon^{\prime} \le 1$.

Fig.~1 is a phase--diagram \cite{pmr} of the system as a function of
$\alpha$ and $\epsilon^{\prime}$. (The regime $\alpha \le 2$ is
featureless and uninteresting.) The different possible dynamical
behavior---periodic, strange nonchaotic and chaotic attractors,
corresponding to the symbols P, S and C in the phase diagram---are
characterised through the largest nonzero Lyapunov exponent,
\begin{equation}
\Lambda= \lim_{N \to \infty}\frac{1}{N}
\sum_{i=1}^N \ln \vert  \alpha(1+\epsilon \cos(2\pi
\phi_i)) (1-2 x_i) \vert.
\end{equation}
We calculate \MLE~ typically from 10$^6$ iterations of the map (after
removing transients for 10$^5$ iterations) which is sufficient to
converge the results to within 10$^{-4}$. To obtain Fig.~1, \MLE~ has
calculated in a 100 $\times $ 100 grid, and the boundary of the chaotic
regions are determined from the $\Lambda = 0$ contour.

The main interesting feature of this phase diagram is the presence of
two separate chaotic regimes, C$_1$ and C$_2$. The C$_1$ region is the
continuation of the chaotic regime in the logistic map (this appears,
for $\epsilon = 0$ at the end of the period--doubling cascade, at
$\alpha = 3.5699\ldots$). The superstable orbit which exists at
$\epsilon = 0, \alpha = 2$ continues (slightly distorted) for nonzero
$\epsilon^{\prime}$ along the locus $\epsilon^{\prime} \approx
2(1-2/\alpha)/5(4/\alpha -1)$; this line (the dotted curve in Fig.~1a)
separates C$_1$ and C$_2$.  The latter chaotic region is one of low
nonlinearity and large--amplitude forcing. The two types of chaotic
attractors that occur in these two regions are qualitatively quite
different---see Fig.~2a and 2b.

SNAs are found in vicinity of the boundaries of the chaotic regions,
where $\Lambda < 0$. The different mechanisms through which they are
formed in different regions are discussed below.

Along the left edge of the region C$_2$, for $0.59 \le
\epsilon^{\prime} \le 1$, the route to SNA is through
fractalization. This region lies to the left of the curve of
``superstable 1-torus'' points (the dashed line in Fig.~1a), namely
well before torus doubling occurs.  For fixed $\epsilon^{\prime}$, SNAs
are created through fractalization in an interval in $\alpha$ as
indicated in the phase diagram.

Along the right edge of C$_2$, SNAs are born through {\it
intermittency}. Consider a superstable 1--torus.  In the regime $0.59
\le \epsilon^{\prime} \le 1$, as $\alpha$ is decreased, holding
$\epsilon^{\prime}$ fixed, a saddle-node bifurcation takes place and
the torus becomes a SNA, eventually becoming a chaotic attractor. This
process occurs over a somewhat narrow range of parameters as compared
to the other routes discussed in Section~1.

Along the left edge of C$_1$ SNAs are created either via the HH
mechanism, or through fractalization. The phenomenon of truncation or
interruption of the period doubling cascade resulting from the
quasiperiodic driving was noted by Kaneko \cite{kk}, who also showed
that there was a power--law scaling between the number of doublings and
the driving amplitude. The requirement, in the Heagy--Hammel mechanism,
that a period doubled torus collide with its unstable parent is a
stringent one, and this is achieved only for selected parameter
intervals. Fractalization is also a likely cause for the interruption
of period doubling \cite{kk}.  The actual mechanism through which a SNA
is created is easily identified from the morphology of the attractor: a
$k$--Torus gives rise to a $k/2$--band SNA in the Heagy--Hammel
mechanism, and to a $k$--band SNA through fractalization. At low
forcing amplitude the fractalization is found to be dominant while HH
mechanism is more common at higher forcing. In particular, we do not
find any fractalization event above $\epsilon^{\prime} \sim 0.7$.

The regions of SNA are quite complicated. The boundaries between
chaotic attractors and SNAs are convoluted, with regions of torus
attractors interspersed among the regions of SNAs. Thus there are
several transformations that the attractors of this system undergo
apart from the birth and death mechanisms, namely the transition from
torus to SNA or SNA to chaos. Indeed, along a line of constant
$\epsilon$ or constant $\epsilon^{\prime}$, \MLE~ is highly
nonmonotonic and reveals a number of bifurcations. An example is shown
in Fig.~3, for constant $\epsilon = 0.05$. Clearly, there are
transitions from T $\to$ SNA $\to$ Chaos $\to$ SNA $\to$ Chaos
$\ldots$, finally terminating in the C$_1$ chaotic region.

Within the SNA region, there are transitions from 2$^n$--band SNAs to
2$^{n-1}$--band SNAs. At these transitions again there are distinctive
and unusual signatures in the dependence of \MLE, as well as in the
distribution of finite--time Lyapunov exponents. This is discussed in
the following section.

Compared to the system at $\epsilon = 0$, for nonzero $\epsilon$ it is
usually possible to observe regular motion only on $n$-T with $n =1, 2,
3, 4, 6, 8, 16$. Higher $n$--frequency tori are surely present, but the
regions where they occur are extremely narrow. The period-3 window of
the logistic map extends as a narrow region of stability wherein one
can locate 3-T, 6-T, and the associated SNAs, which are formed through
the same general mechanisms as outlined in Section~I. A blow--up of
this region of the phase diagram (marked by W in Fig.~1a) is shown in
Fig.~1b.  The intermittent route to SNA that operates on the right edge
of C$_2$ occurs here along the top edge of the periodic 3-T region.  The
3--band and 6--band SNAs that occur appear to be formed through the
fractalization or Heagy-Hammel route.

This qualitative picture appears to be valid for all quasiperiodically
forced unimodal maps. We have studied the additively forced logistic
map, as well as the forced sine--map and found a similar phase diagram.
For higher dimensional systems, the picture gets somewhat more
complicated, but the essential qualitative features carry over, much in
the same manner as the bifurcation diagram for 1--$d$ maps generalizes
for higher dimensional dissipative chaotic systems. This system is also
not very sensitive to the numerical value of the irrational driving
frequency $\omega$: qualitatively similar phase diagrams have been
obtained for other $\omega$'s.

\section{Results and Discussion}
In this section we study the variation of the dynamics through several
transitions in the system as the parameters $\alpha$ and $\epsilon$ are
varied. In addition to the Lyapunov exponent \MLE~ itself, we also
examine the $N$--step Lyapunov exponents, $\lambda_N$, namely
\begin{equation}
\lambda_N = \frac{1}{N} \sum_{i=1}^N \ln \vert \alpha(1+\epsilon
\cos(2\pi \phi_i)) (1-2 x_i) \vert, 
\end{equation}
their variance, $\sigma$, and the distribution
\beq P(N,\lambda) d\lambda = {\rm Probablity~ }~{\rm that~} \lambda_N
{\rm ~lies~ between}~ \lambda ~ {\rm and}~ \lambda + d\lambda.
\eqn
This is of great relevance when studying the stability of systems where
a small change in control parameter gives raise to drastic change in
dynamical behavior. In our calculations, \MLE~ and its variance are
typically computed from a sample of 50 estimations of step length $N =
10^5$.

The SNA $\to$ chaos route, which has been studied extensively and where
it is known that the Lyapunov exponent varies linearly through the
transition \cite{lai}, will not be considered here.

\subsection{From Tori to SNAs}

\subsubsection{The Heagy--Hammel mechanism}

In the HH mechanism \cite{hh}, a period $2^n$ torus attractor gets
wrinkled and upon collision with the parent unstable period $2^{n-1}$
torus, a $2^{n-1}$ band SNA is formed. This phenomenon is similar (in
some sense) to the band--merging crisis that occurs in unforced
systems, where the Lyapunov exponent is known to have a discontinuous
slope \cite{mr}. In contrast, in Ref.~\cite{hh} the transition to SNA
appeared to be smooth as a function of the parameter (see Figs.~2 and
10 in \cite{hh}).

We examine this behaviour in some detail in Fig.~4a, where for
$\epsilon^{\prime} =0.3$ the crisis takes place at $\alpha_{HH} =
3.487793 \ldots$. When examined in a sufficiently small neighborhood of
$\alpha_c$, the transition is clearly revealed by \MLE: on the torus,
\MLE~ varies smoothly but in the SNA phase the variation is
rather irregular and the crossover between these two behaviours is
abrupt.  It is also possible to identify the transition point from the
examination of the variance in \MLE~, shown in Fig.~4b: in the torus
region, $\alpha \le \alpha_{HH}$, the fluctuations in \MLE~ are small,
while for $\alpha > \alpha_{HH}$, the the variance is large and depends
irregularly on the function of the control parameter.  Unlike the case
of band--merging or widening crises \cite{mr} there is no distinctive
signature in the variation of the Lyapunov exponent itself (except for
a certain irregularity in the SNA phase).

\subsubsection{The Fractalization mechanism}
During fractalization, a period $k$ torus attractor get wrinkled and
eventually forms a $k$ band SNA\cite{lai}. There is no apparent
interaction with unstable periodic orbits in contrast to the the HH
case, and there no analogue of any crisis--like behavior. The variation
of the \MLE~ and its variance at such a transition are shown in
Figs.~5a and 5b respectively, as $\alpha$ changes through $\alpha_F =
2.6526 \ldots $ at $\epsilon^{\prime} = 1$. The behavior of variance in
Fig.~5b is similar to that of the HH case (Fig.~4b), except in the
magnitude of the fluctuations.  These figures also show that the
transition from torus to SNA is smooth with no particular signature in
\MLE.

\subsubsection{The Intermittency route}
Intermittent SNAs are morphologically quite distinct from those formed
through other mechanisms. An example of the transition to such SNAs is
shown in Fig.~6. The behavior of \MLE~ as $\alpha$ is reduced through
$\alpha_I = 3.405808806 \ldots$ at $\epsilon^{\prime} = 1$ (Fig.~6a) is
distinctive and may be contrasted with Figs.~4a and 5a.  The torus
which exists for $\alpha > \alpha_I$ and the intermittent SNA are
depicted in Figs.~6b and 6c respectively. On the intermittent SNA, most
points remain near the parent torus, with sporadic large deviations.
On fractalized SNAs points on the attractor stay close to the parent
fractal torus (Fig.~1 of Ref. \cite{nk}), while on HH SNAs, points are
distributed within the entire region enclosed by the wrinkled bounding
tori (see Fig.~1 of Ref. \cite{hh}). The other characteristic behavior
can be extracted from Fig.~6d where we plot the variance: this changes
abruptly at the transition, where a saddle-node bifurcation occurs
showing the characteristic signature of the intermittency route to
SNAs.

\subsubsection{Finite--time Lyapunov exponents}

As is well known, while \MLE~ is negative on a SNA, for short times,
the local Lyapunov exponent can be positive. One of the characteristics
of the SNAs born through different mechanisms is the difference in the
distribution of finite-time exponents, namely $P(N,\lambda)$. In the
limit of large $N$, it is clear that this distribution will collapse to
a delta-function, $\lim_{N \to \infty} P(N,\lambda) \to
\delta(\Lambda-\lambda)$.
The deviations from---and the approach to---the limit can be very
different for SNAs created through different mechanisms. 

Shown in Figs.~7a--7c are the distributions for $P(50,\lambda)$ across
the three transitions discussed above, namely on the tori and
corresponding SNAs. A common feature of all three cases is that
$P(N,\lambda)$ is strongly peaked about \MLE~ when the attractor is a
torus, but on the SNA, the distribution picks up a tail which extends
into the $\lambda >0 $ region. This tail directly correlates with the
enhanced fluctuation in \MLE~ on SNAs (see Fig.~4b, 5b or 6d).  On the
fractalized SNA, the distribution shifts continuously to larger \MLE,
but the shape remains the same for $\alpha < \alpha_I$ and $\alpha >
\alpha_I$, while on the HH or intermittent SNA, the actual shape of
the distribution on the torus and the SNA are very different.

One remarkable feature of intermittent SNAs is that the postive tail in
the distribution decays very slowly: even for $N$ as large as 10$^4$,
it does not completely disappear\cite{pmr}. In order to quantify this,
we define the fraction of positive local Lyapunov exponents as
\beq
F_+(N) = \int_0^{\infty} ~P(N,\lambda)~ d\lambda,
\eqn
and similarly, the first moment, namely a local K--S entropy,
\beq
K(N) = \int_0^{\infty} \lambda~P(N,\lambda)~ d\lambda
\eqn
Clearly, $\lim_{N \to \infty} F_+(N) \to 0$ and $\lim_{N \to \infty}
K(N) \to 0$. Empirically we have found that on the intermittent SNA,
these quantities show the large $N$ behavior
\beq
F_+(N) \sim N^{-\beta}
\eqn
(similarly for $K(N)$) while for the fractalized or HH SNAs, the
approach is exponentially fast,
\beq
F_+(N) \sim \exp(-\gamma N).
\eqn
The exponents $\beta$ and $\gamma$ depend strongly on the parameters
$\alpha$ and $\epsilon^{\prime}$. For the SNAs discussed above, we have
calculated $F_+(N)$ for large $N$ (see Fig.~7d) and obtain $\beta
\approx 0.2$, $\gamma_{HH} \approx 0.02$ and $\gamma_I \approx 0.03$.
$K(N)$ also has a similar slow fall--off at large $N$ for the
intermittent SNA $K(N) \sim N^{-\beta^{\prime}}$ with $\beta^{\prime}
\approx 0.55$, while for the other SNAs, the decay is slow for small
$N$ but exponential for large $N$.

\subsection{Scaling at the Intermittency transition}

The intermittency transition from a torus to a SNA is characterised by
scaling behavior for the dynamics, similar to corresponding behavior in
the unforced case \cite{cfh}. The `laminar' phase in this case is the
torus, while the `chaotic' phase is the nonchaotic attractor. In order
to obtain the distribution time in these phases, we co-evolve two
trajectories with identical $(x_0,\phi_0)$ and $\epsilon^{\prime}$,
with different $\alpha$: since the angular coordinate remains
identical, the distance between the trajectories is simply the
difference in the $x_n$'s. We calculate the time between bursts and fit
to the scaling form
\beq
\tau \sim (\alpha_c -\alpha )^{-\theta}.
\label{thetaeq} 
\eqn
The numerical value obtained for the attractor with $\alpha$ near
$\alpha_I$ and $\epsilon^{\prime}$ = 1 is $\theta = 0.52 \pm 0.03$ and
$\epsilon^{\prime}$= 0.65 is $\theta = 0.5 \pm 0.03$ (See Fig.~8a).
This suggests that the intermittency is Type-{\rm 1}. 

The Lyapunov exponent itself shows the scaling form \cite{pmr}
\beq
\Lambda-\Lambda_c \sim (\alpha_c -\alpha)^{\mu}, 
\label{inter}
\eqn
at fixed $\epsilon^{\prime}$ which can be compared with the probability
density \cite{mr,pl,goy} in the SNA burst phase (Fig.~8(b)). Both these
quantities have the same exponent $\mu = 0.37\pm0.03$ at
$\epsilon^{\prime} = 1$. Although other SNAs also show
intermittency--like dynamical behavior, the scaling form
Eq.~(\ref{inter}) obtains {\em only} for the intermittent SNAs, thus
providing a means of distinguishing these from HH or fractalized SNAs.

\subsection{Merging of SNAs}

As the parameters $\alpha$ and $\epsilon$ are varied, the attractors of
the quasiperiodically driven system undergo transformation in a manner
which is analogus to the undriven chaotic system. Similar to reverse
bifurcations or band merging in 1--d maps, $n$--band SNAs transform to
$n/2$--band SNAs. Through such a transition, when the dynamics remains
nonchaotic and strange, \MLE~ is a good order parameter.  Sosnotseva
\etl \cite{sfkp}, who discovered the first example of this transition in the
driven H\'enon and circle maps, demonstrated the merging by examining
the phase portait.

Given the fairly narrow range over which SNAs exist in any system, this
transition also occurs in a restricted range. We find that the SNA
bands that are formed by the HH mechanism typically do not merge at
negative Lyapunov exponent, but only do so at higher driving, when they
collide in the chaotic region as at a proper analogue of the band
merging crisis \cite{gory}. However, the variation of \MLE~ at such
transitions does not follow a uniform pattern as in the unforced case
\cite{mr,cfh}. 

Those SNAs which are formed {\it via} fractalization may merge at
negative \MLE. Fig.~9a shows the variation of \MLE~ for such an example
of a 2--band SNA merging to a 1--band SNA.  The Lyapunov exponent {\it
{decreases}} with increasing nonlinearity.  The distribution for short
time Lyapunov exponents before and after the SNA band merging crisis
confirms that the distribution shifts to lower $\lambda$; see Fig.~9b.
Both these figures indicate that the chaoticity of the system (measured
either through local ($\lambda_N$) or global ($\Lambda$) indicators)
becomes significantly lower after merging crisis.  This band merging is
different from unforced case \cite{ott,cfh} where \MLE~ generally
increases as number of bands decreases with increasing nonlinearity.

\subsection{The Effect of Noise}

An important consideration in the study of SNAs is their robustness.
Given the somewhat unusual properties of such attractors and the fact
that they exist over small regions in parameter space, it is natural to
examine the effect of fluctuations. This is of particular relevance
with respect to the experimental observation of SNAs
\cite{ditto,newexp}. The effect of noise in the logistic map has been
extensively studied \cite{cfh}, and it is known that noise generally
lowers the threshold for chaos --- systems with additive noise have a
larger Lyapunov exponent for smaller nonlinearity. Furthermore,
transitions and bifurcations get ``blurred''in the presence of
fluctuations.

To examine some of these effects, we have studied this system through
the introduction of additive noise in the dynamics, for example as
\begin{equation}
x_{n+1} = \alpha (1 + \epsilon \cos (2 \pi \phi_n))~x_n~ ( 1 - x_n) +
\rho \xi_n
\eqn
where $\rho$ is the noise amplitude and the random variable $\{\xi\}$
is $\delta$--correlated in time. As may have been anticipated, the
addition of noise ``smears out'' tori, and the threshold values for
bifurcations typically shift to lower $\alpha$ at fixed $\epsilon$.

The actual transitions---now from noisy tori to noisy SNAs---survive,
and examples of this are shown in Fig.~10 for fractalized, HH and
intermittent SNAs, where \MLE~ is shown as a function of $\alpha$ with
and without noise. Since as a function of decreasing $\alpha$ the
intermittent SNA is born out of a torus, the effect of noise
is to increase the parameter value at which this SNA is created
(relative to the $\rho = 0$ case) while the opposite effect, namely a
reduction of parameter value for the transition to HH and fractalized
SNAs is seen. These behaviors are typical: we have verified that SNAs
in related systems such as the quasiperiodically forced ring and
H\'enon maps behave similarly.  

Upon addition of noise, the global structure of the different types of
SNAs remain similar---the qualitative behavior such as scaling, band
merging \etc are retained---although the numerical values of constants
and exponents change. The degree of robustness varies with
nonlinearity, though: for instance, at $\epsilon^{\prime} = 1$, the
intermittent transition survives for additive noise of amplitude up to
$\rho = 10^{-6}$, while at lower $\epsilon^{\prime}$ the transition is
robust to even larger $\rho$ $\approx 10^{-4}$. The region of scaling
in presence of noise is small, and the intermittency exponent varies;
we have found that the numerical value of $\theta$ (see
Eq.~(\ref{thetaeq})) is $ 0.56 \pm 0.04$ at $\epsilon^{\prime} = 1$.

\section{Summary}

In this paper we have described the phenomenology of strange nonchaotic
dynamics in a prototypical example, namely the quasiperiodically driven
logistic map. There are three different mechanisms through which SNAs
can be created in this system; these routes to SNAs have their
analogues in the different scenarios for the onset of chaos in
dissipative dynamical systems \cite{ott}.

We obtain a ``phase--diagram'' for the system, delineating the
different asymptotic behavior that are possible as a function of the
parameters, namely $n$--frequency torus attractors, strange chaotic
attractors and SNAs. There are two major chaotic regions with different
characteristics, separated from each other by a region of quasiperiodic
and strange nonchaotic behavior.  As a function of parameters,
therefore, the system can show several transitions in the dynamics and
several of these have been studied.

To distinguish among the different mechanisms though which SNAs are
born, we examine not only the manner in which \MLE~ changes as a
function of parameters, but also the variance, $\sigma$.  These indices
together give a clear indication of the transition from quasiperiodic
to strange nonchaotic dynamics. We also examine the distribution of
local Lyapunov exponents, $P(N,\lambda)$ and find that on different
SNAs, the fraction of positive exponents $F_+(N)$ or its moment, the
local KS entropy, decay in different manners depending on the type of
SNA that is formed.

In addition to three routes that have been decribed previously
\cite{hh,k,yl}, we identify a new mechanism for the creation of
SNAs. These {\it intermittent} SNAs \cite{pmr} are atypical in that
they are created at the quasiperiodic analogue of a saddle--node
bifurcation \cite{ott}, and the signature of the transition is a
discontinuous change both in \MLE~ as well as in $\sigma$. The chaotic
component on the intermittent SNA is long lived, giving rise to long
chaotic transients: this shows up as a slowly decaying positive tail in
$P(N, \lambda)$, and a resulting power--law decay for $ F_+(N)$ or
K(N). (On other SNAs, in contrast, these quantities decay
exponentially.) We further characterise intermittent SNAs by
establishing the scaling behaviour for residence times in the different
phases and find that the qualitative picture is in accord with Type--I
intermittency \cite{ott}.  This behaviour persists in the presence of
additive random noise.

We have described the intermittent SNA in detail and shown that both in
its creation as well in its morphology, it is distinct from other SNAs,
and bears some relation, in terms of the phase diagram, to a
re--entrant phase. (In the higher dimensional system of the forced
circle map \cite{sfkp} for which the phase diagram has been obtained,
the intermittent SNA occurs in an analogus region). In the vicinity of
such attractors, a number of dynamical quantities show scaling
behaviour.

There are other bifurcation phenomena in such systems, and we have
examined the case of SNA band merging, namely the coalescence of
different branches of a multiband attractor. At this transition, in
contrast to analogous behaviour in the logistic map \cite{mr}, \MLE,
which remains negative throughout, decreases. A clear understanding of
why this is so is not available at present.

In summary, our primary emphasis has been in the use of a number of
Lyapunov measures---the largest nontrivial Lyapunov exponent, its
fluctuations, the distribution of finite time Lyapunov exponents and
partial moments of this distribution---to characterise the different
types of transitions to strange nonchaotic attractors that arise in the
forced logistic map.

For hyperbolic systems, the theory for the Lyapunov exponent and for
Lyapunov measures is well developed \cite{grass}. For SNAs, the
situation is in a less satisfactory state, and our work represents
initial efforts towards understanding the phenomenology of
quasiperiodically driven systems. For a number of bifurcations in such
systems in general, it is clear that the largest Lyapunov exponent is a
good order--parameter. It is likely that the considerable formalism for
such transitions that has been developed for chaotic strange attractors
\cite{ott,grass} can be applied in large part to the strange nonchaotic
regime, but the extent to which the theory carries over is an aspect
that remains to be explored in future work.

\vskip1cm
{\sc ACKNOWLEDGMENT}: This research was supported by grant SPS/MO-5/92
from the Department of Science and Technology, India.

\newpage
\centerline{Figure Captions}

\begin{itemize}
\item[Fig. 1]
(a) Phase diagram for the forced logistic map (schematic). The rescaled
parameter $\epsilon'$ is defined as $\epsilon'=\epsilon/(4/\alpha-1)$.
T and C correspond to Torus and Chaotic attractors. The shaded region
along the boundary of T and C corresponds to SNA (marked S).  The
boundaries separating the different regions are convoluted, and regions
of SNA and Chaotic attractors are interwoven in a complicated manner.
The dashed curve marks the locus of the ``superstable'' orbit (see the
text). The region of periodic attractors can be further demarcated into
period 1, 2 and 4 tori as shown (1T, 2T and 4T).  W denotes the window
of periodic behavior corresponding to the period--3 orbit of the
logistic map. This is shown enlarged in Fig.~1b.  Intermittent SNAs are
found on the edge of the C$_2$ region marked I, while the left boundary
of C$_2$ has only fractalised SNAs. Along the boundary of C$_1$, both
fractalised and HH SNAs can be found.\\ (b) A blow-up of the window W
indicated in Fig~1(a).  This small window shows periodic tori of period
3,6,.. and their SNAs that are created either through the Heagy-Hammel
process or fractalization. As may be expected, other similar windows
can be seen upon further enlargement.

\item[Fig. 2] Typical chaotic attractors in the two regions
a) C$_1$, for $\alpha=3.6$, $\epsilon^{\prime}= 0.5$, and 
b) C$_2$ at $\alpha=3$ $\epsilon^{\prime}= 1$.

\item[Fig. 3] Variation of \MLE~ as a function of  $\alpha$ for fixed
$\epsilon = 0.05$. Note the highly oscillatory structure indicative of
several transitions in the system.

\item[Fig. 4] The transition from a period 2--torus to 1--band SNA though
the Heagy-Hammel mechanism along the line $\epsilon^{\prime} = 0.3$ and
at $\alpha_{HH} = 3.487793\ldots$.  a) behavior of \MLE~through the
transition,  and b) the variance.

\item[Fig. 5] The transition from a period 1-torus to 1--band SNA via
the fractalization route along the line $\epsilon^{\prime} = 1$ and at
$\alpha_F = 2.6526\ldots$.    a) behavior of \MLE~through the
transition, and  b) the variance.

\item[Fig. 6] The transition from a period 1-torus to 1--band SNA via
the intermittent transition, along the line $\epsilon^{\prime} = 1$ and
at $\alpha_I = 3.405808806\ldots$. a) behavior of \MLE~through the
transition, b) Plot of the period 1-Torus at $\alpha=3.405809$, c) the
intermittent SNA at $\alpha=3.405808$, d) the variance through the
transition.

\item[Fig. 7]
Distribution of finite time Lyapunov exponents across the three
transitions. Shown are the distributions of $P(50,\lambda)$ before and
after the transitions, namely on the tori and on the SNA. a) Along the
HH route, on the torus at $\alpha=3.4874$ ($\Box$) and on the SNA at
$\alpha =3.488$ ($\triangle$) for $\epsilon^{\prime}= 0.3 $. b) Along
the fractalization route, on the torus at $\alpha=2.63$ ($\Box$) and
the SNA at $\alpha =2.66$ ($\triangle $) for $\epsilon^{\prime}=1$. c)
Along the intermittency route, on the torus at $\alpha=3.40581$
($\Box$) and on the SNA at $\alpha =3.4058056$ ($\triangle$)  for
$\epsilon^{\prime}=1$. d) Variation of $F_+(N)$ for the three different
SNAs in a), b) and c) showing exponential decay in the first two cases,
and a power--law decay for the intermittent SNA. The respective symbols
are $\Box$,$\circ$ and $\nabla$.

\item[Fig. 8] Scaling behavior on the intermittent SNA.
a) Plot of the average time between bursts {\it vs} $(\alpha_c-\alpha)$
at $\epsilon^{\prime} = 1 (\circ)$ and at $\epsilon^{\prime} = 0.65
(\Box)$.  (b) The probability density ($p_B$) in the burst phase
($\circ$) and $\Lambda$ ($\Box$) {\it vs} $(\alpha_c-\alpha)$ at
$\epsilon^{\prime} = 1$. The measured exponents are $\mu \approx 0.37$.

\item[Fig. 9] a) Variation of \MLE~ at band--merging bifurcation in SNA
for 2--band SNAs to a 1--band SNA along $\epsilon=0.05$.  b)
Distribution of local Lyapunov exponents, $P(100,\lambda)$, across the
merging transition in a) which takes place at $\alpha_c
\approx 3.387439$. 

\item[Fig. 10] The shift in \MLE~ with additive noise near the
threshold for the transition from SNA to chaos for the cases of a) HH,
b) Fractalization, for $\rho=0$ (solid line), $10^{-3}$ (dashed line)
and c) Intermittent SNAs, $\rho = 0$ (solid line) $\rho =10^{-5}$
(dashed). All quantitites are estimated from a set of 50 samples of
10$^5$ time-steps.

\end{itemize}

\begin{thebibliography}{99}
\bibitem{ring}
This route is quite general and occurs in higher dimensional systems as
well. In the quasiperiodically forced ring map \cite{sfkp} (see Fig.~3
there for a phase diagram) intermittent SNAs occur (for example at $k =
3$ and $ A = 0.1851469$), and are apparently created by a similar
saddle--node bifurcation.
\end{thebibliography}
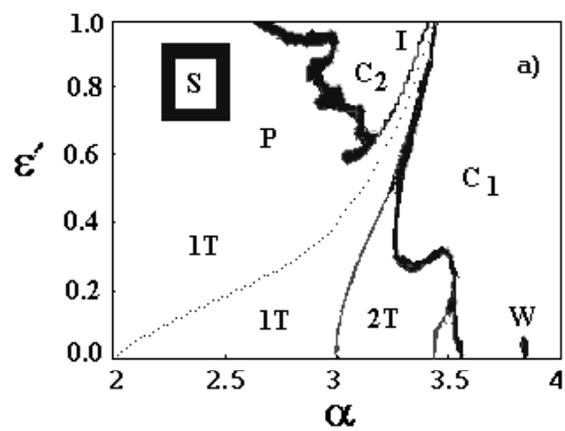
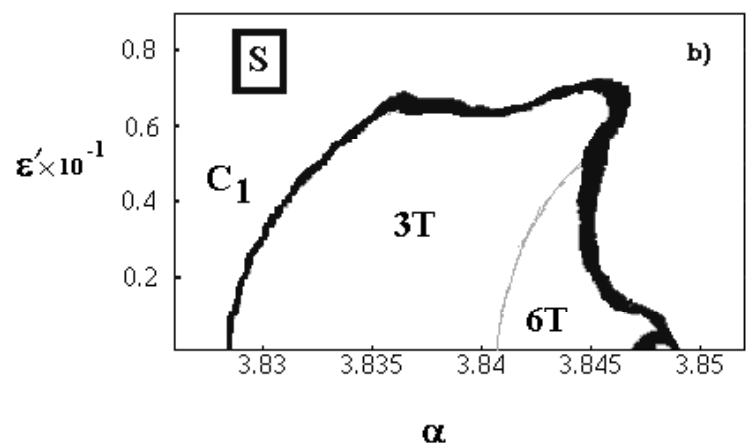
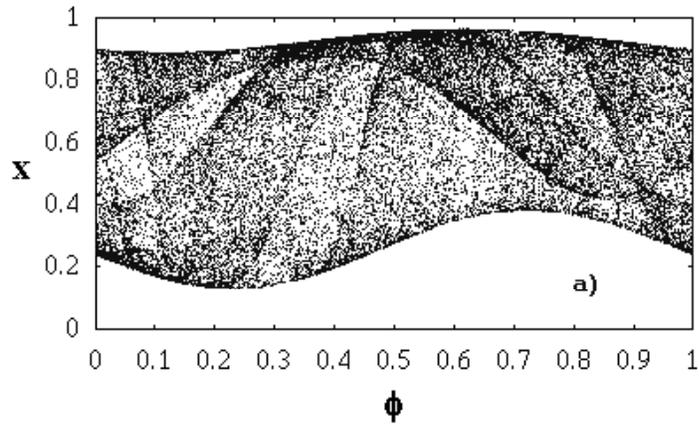
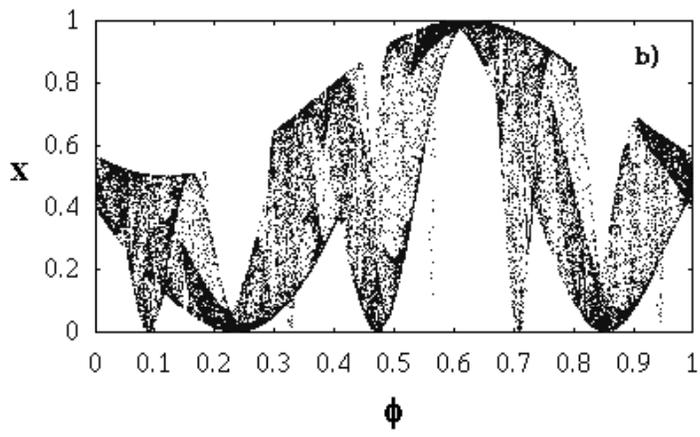
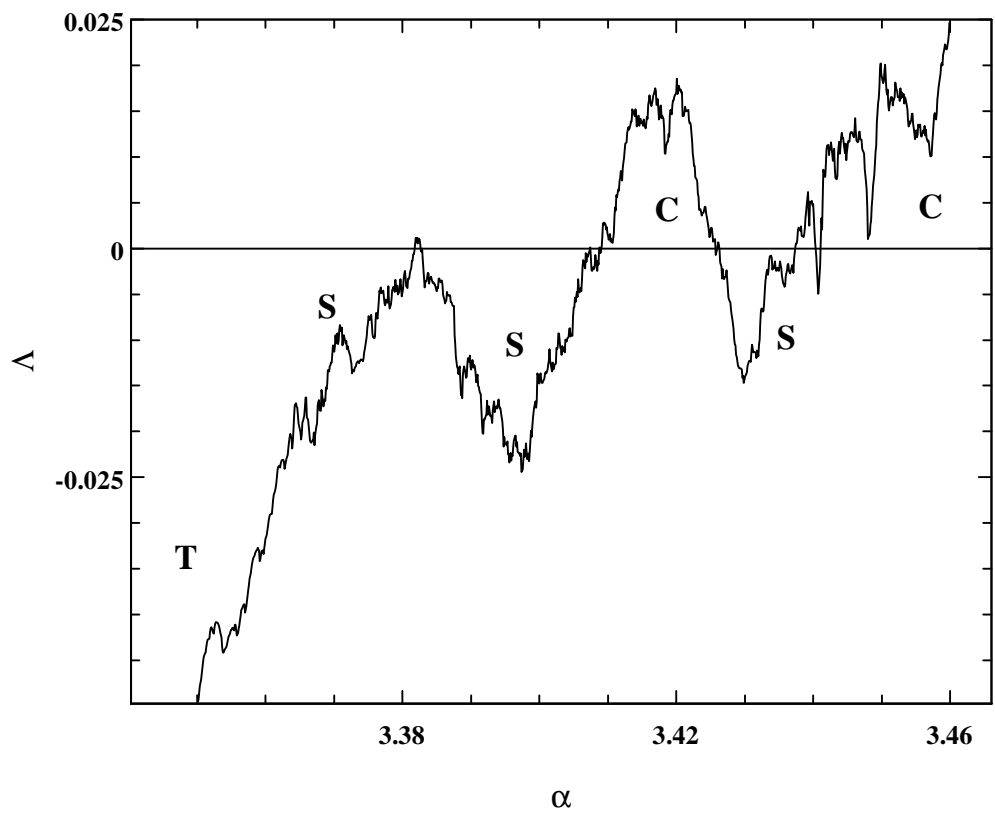
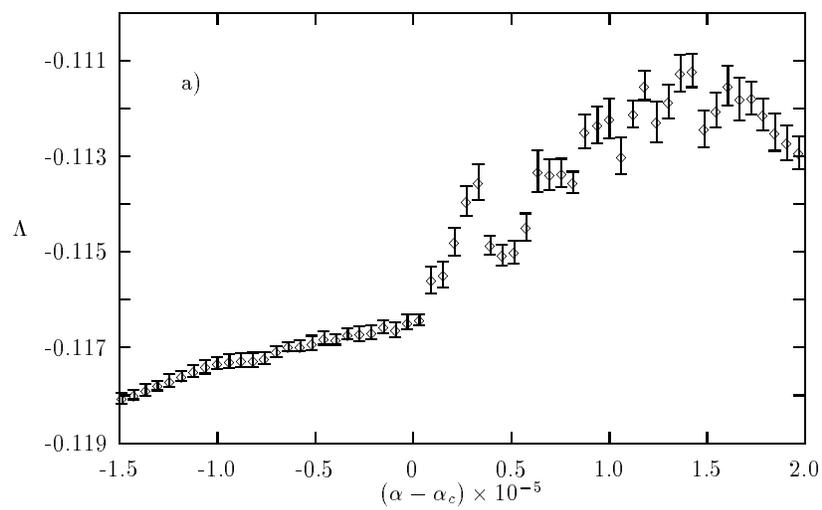
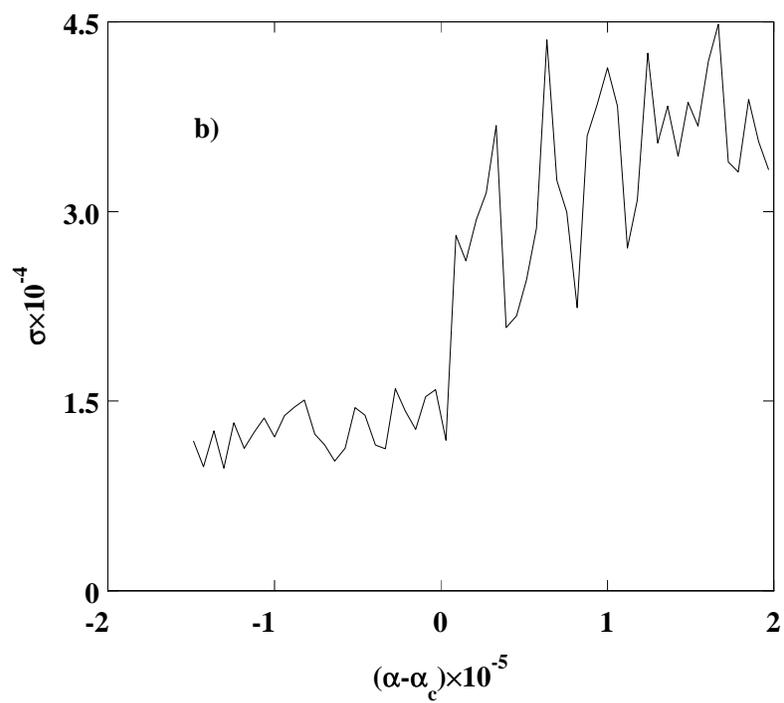
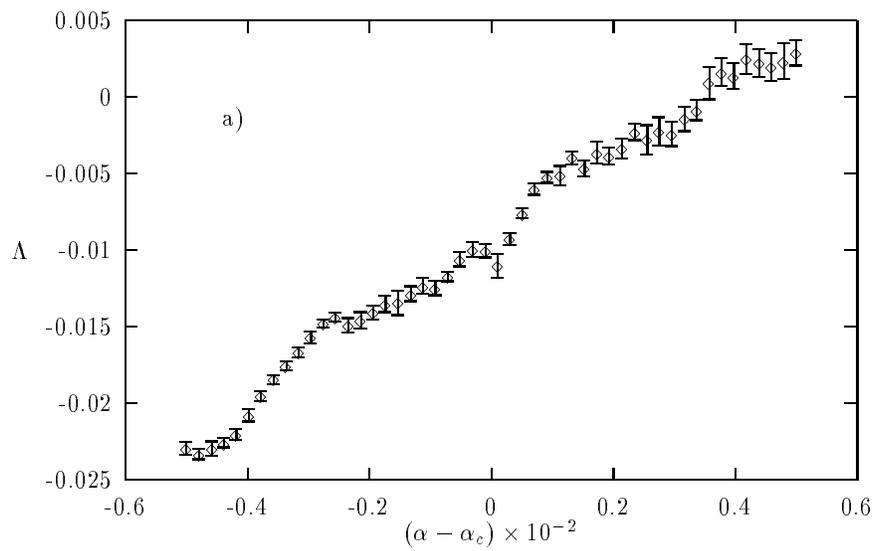
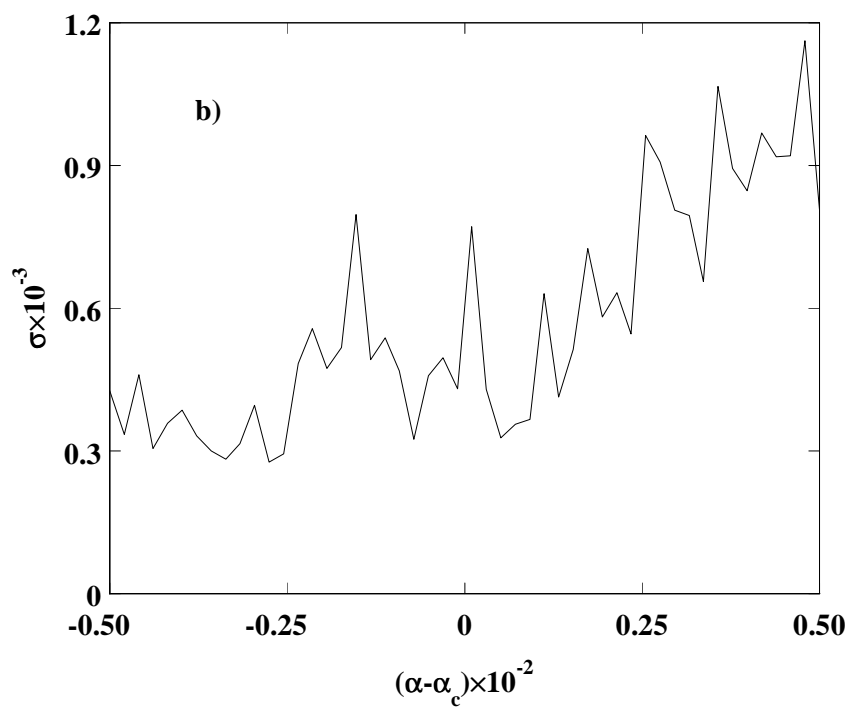
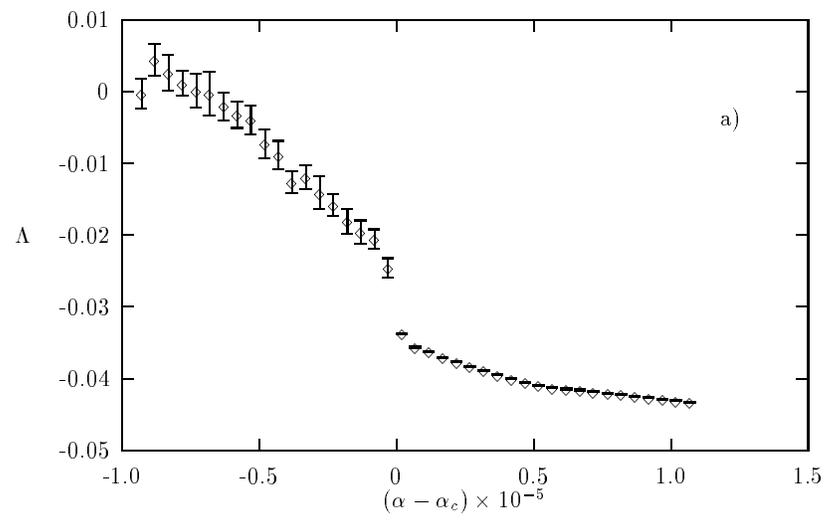
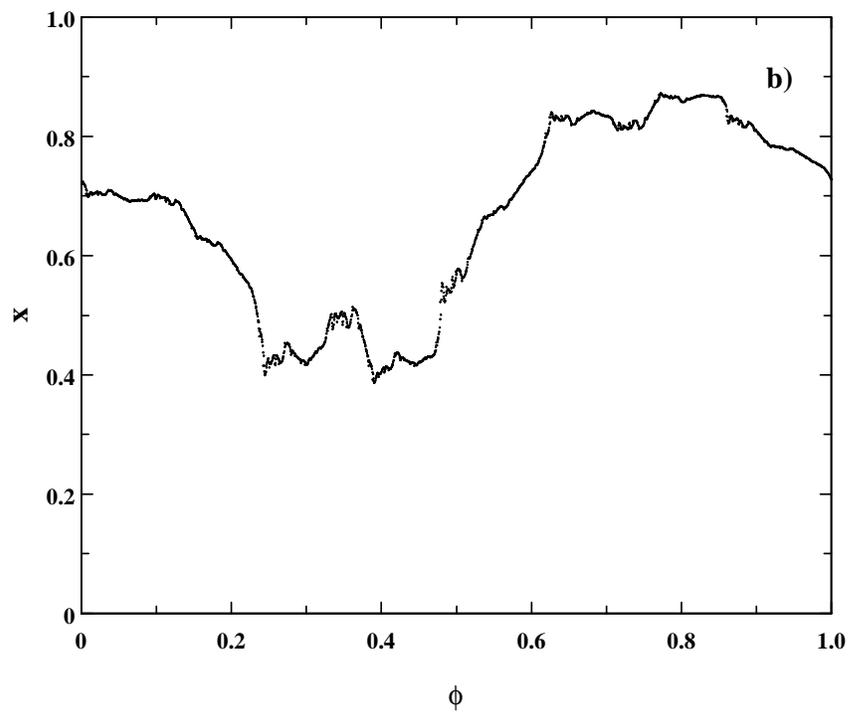
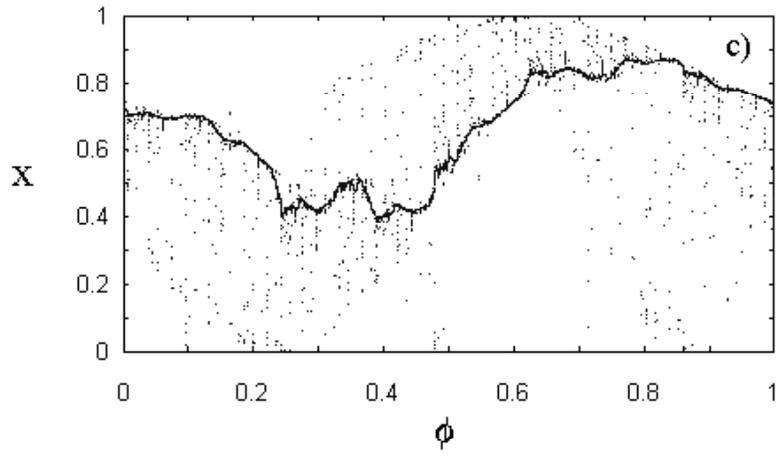
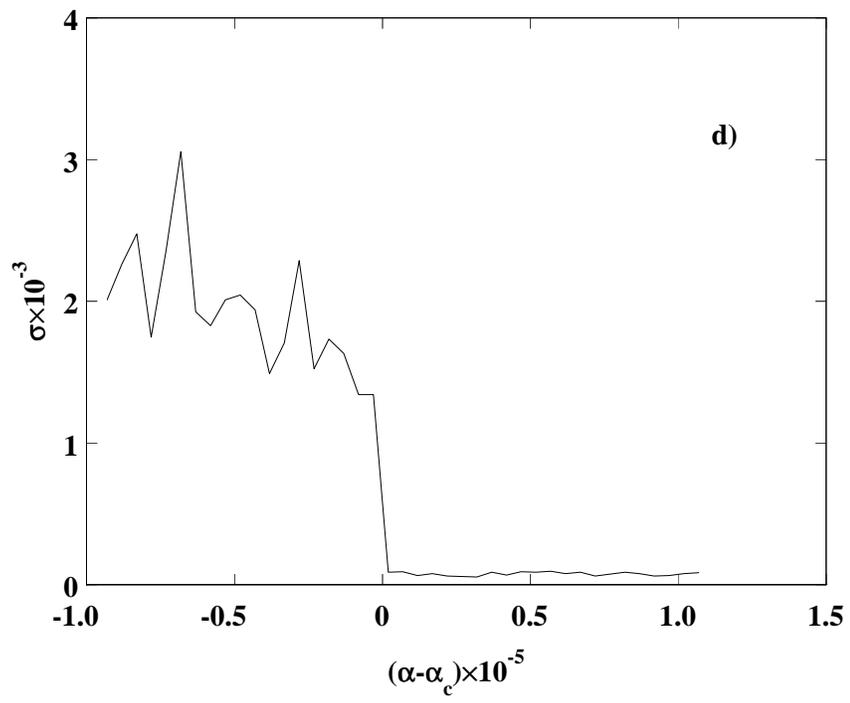
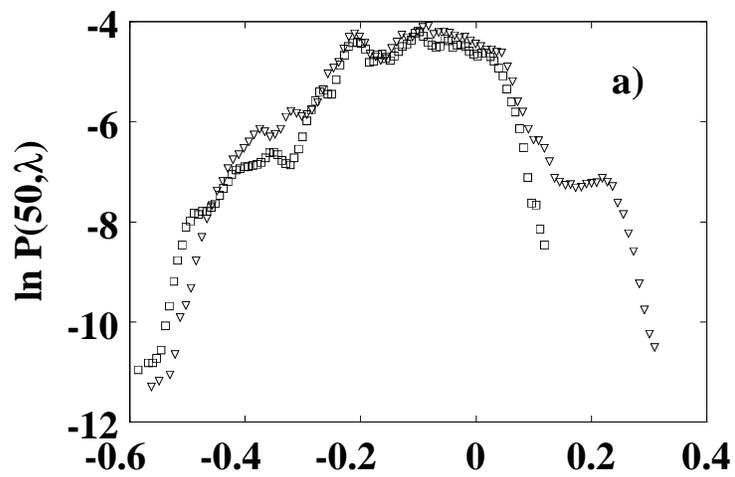
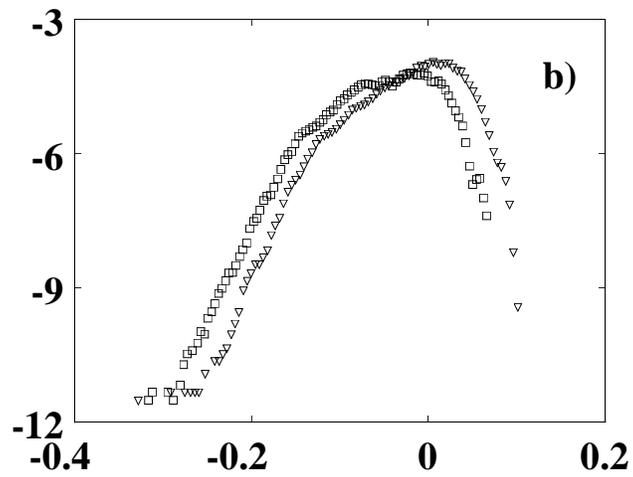
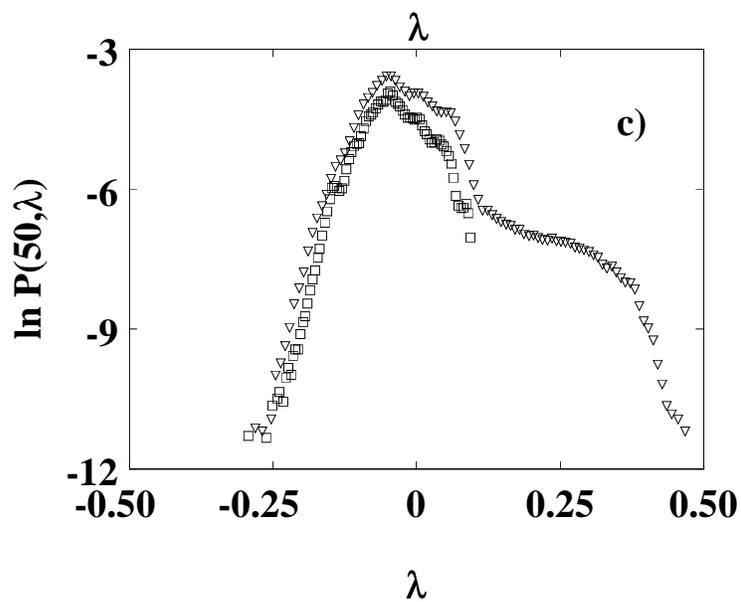
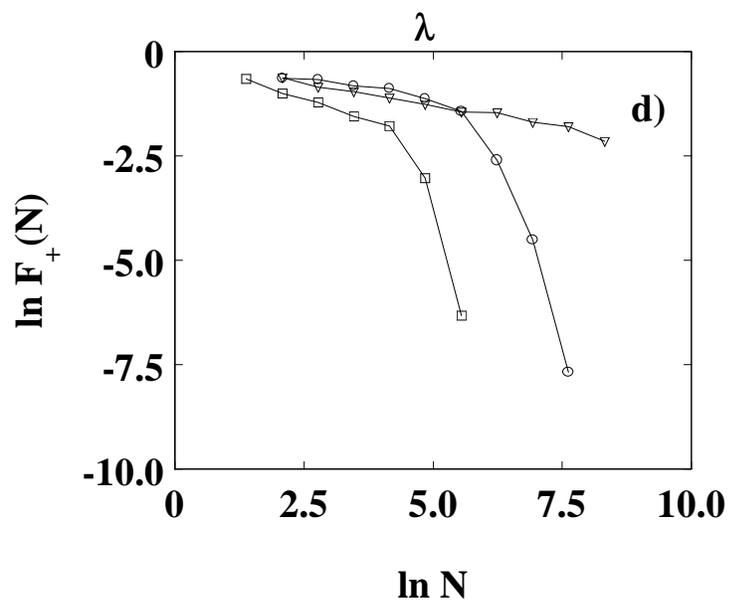
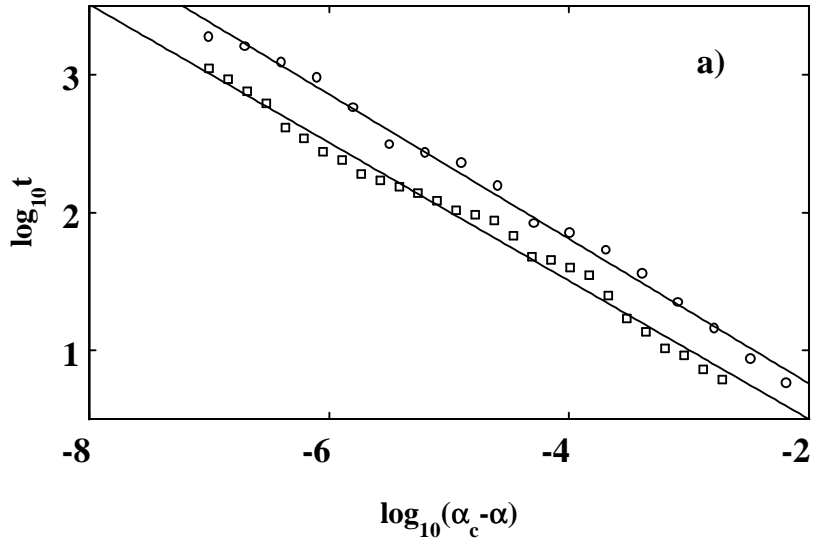
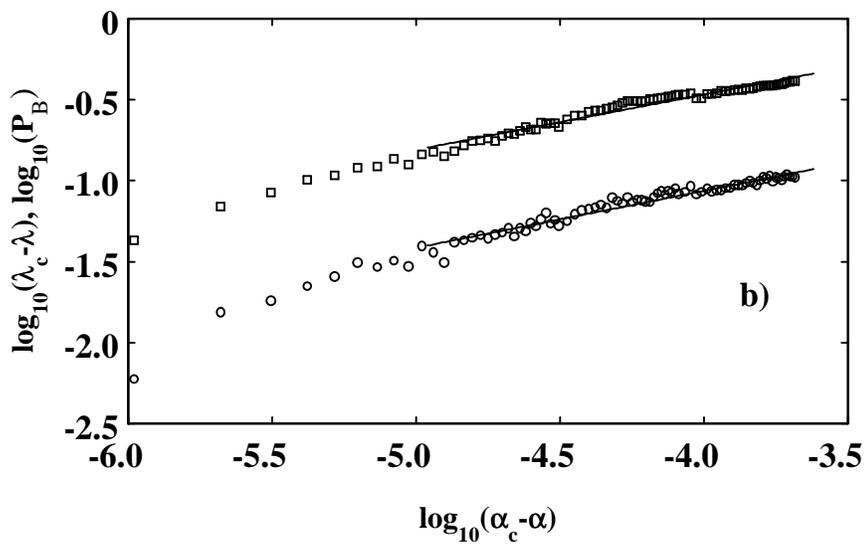
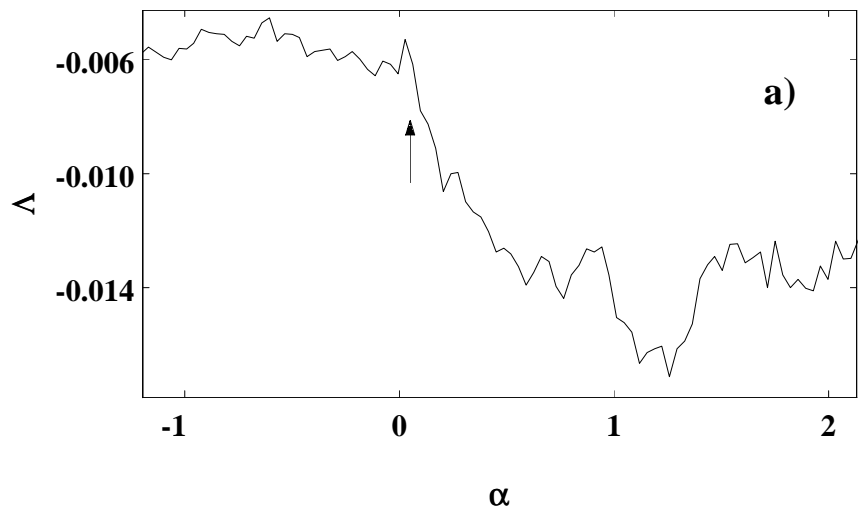
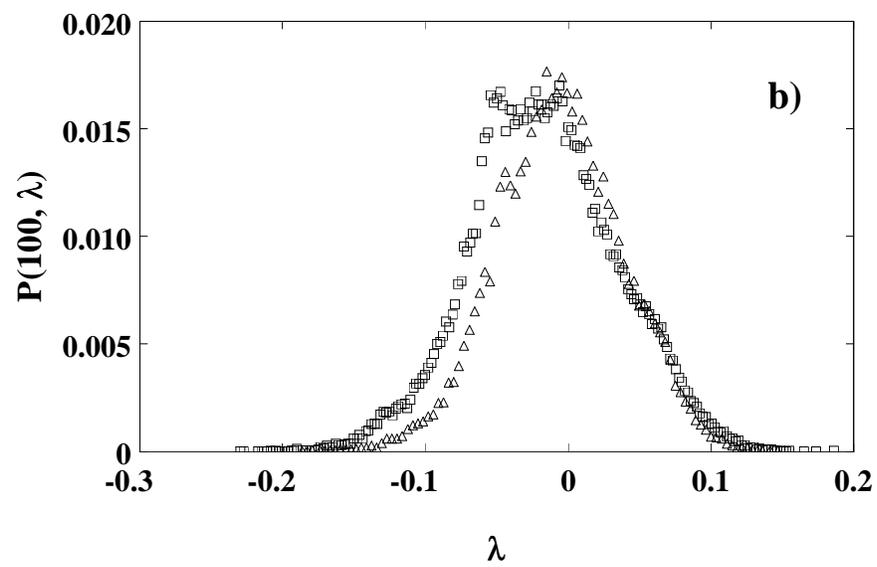
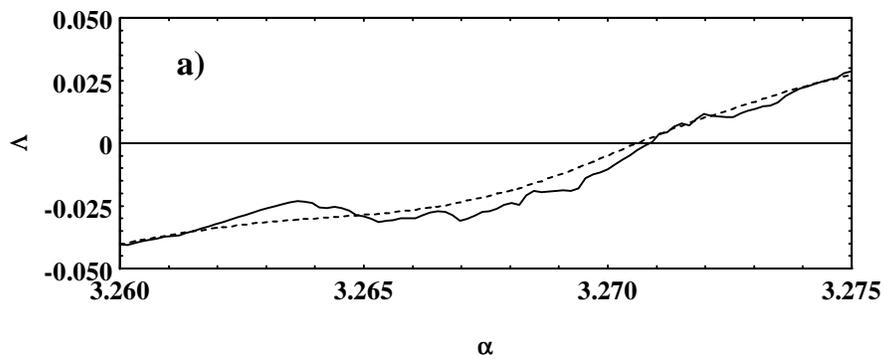
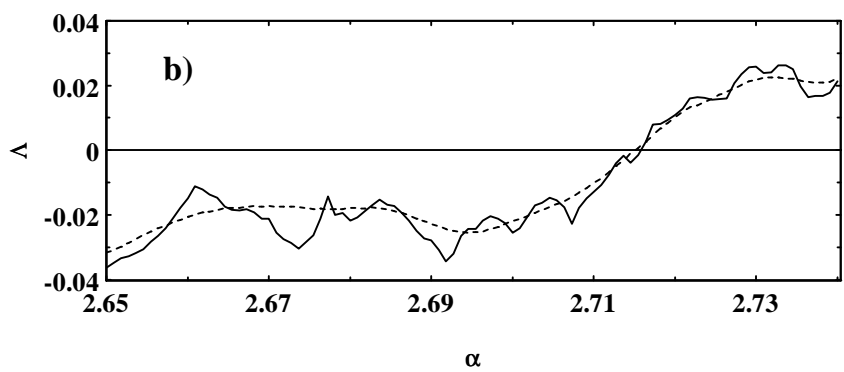
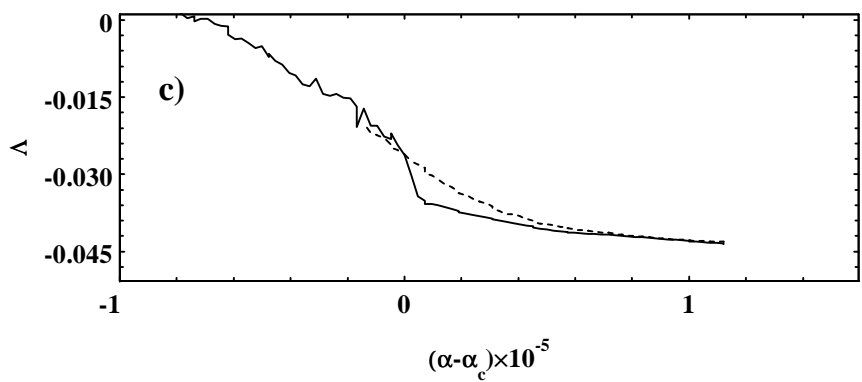
\end{document}